\documentclass[aps,prl,unsortedaddress,twocolumn,uperscriptaddress]{revtex4-1}

\usepackage{amsmath}
\usepackage{graphicx}

\graphicspath{{./}{figures/}}

\begin{document}
\title{Coherent order parameter dynamics in SmTe$_3$} 

\author{M. Trigo}
\email[E-mail: ]{mtrigo@slac.stanford.edu}
\affiliation{Stanford PULSE Institute, SLAC National Accelerator Laboratory, Menlo Park, CA 94025, USA}
\affiliation{Stanford Institute for Materials and Energy Sciences, SLAC National Accelerator Laboratory, Menlo Park, CA 94025, USA}

\author{P. Giraldo-Gallo}
\affiliation{Department of Applied Physics, Stanford University, Stanford, CA 94305, USA}
\affiliation{Department of Physics, Universidad de Los Andes, Bogot\'a, 111711, Colombia.}

\author{M. E. Kozina}
\author{T. Henighan}
\author{M. P. Jiang}
\author{H. Liu}   
\affiliation{Stanford PULSE Institute, SLAC National Accelerator Laboratory, Menlo Park, CA 94025, USA}
\affiliation{Stanford Institute for Materials and Energy Sciences, SLAC National Accelerator Laboratory, Menlo Park, CA 94025, USA}
\author{J. N. Clark}
\affiliation{Stanford PULSE Institute, SLAC National Accelerator Laboratory, Menlo Park, CA 94025, USA}

\author{M. Chollet}
\author{J. M. Glownia}
\author{D. Zhu}
\affiliation{Linac Coherent Light Source, SLAC National Accelerator Laboratory, Menlo Park, California 94025, USA}

\author{T. Katayama}
\affiliation{Japan Synchrotron Radiation Research Institute, 1-1-1 Kouto, Sayo-cho, Sayo-gun, Hyogo 679-5198, Japan}

\author{D. Leuenberger}
\affiliation{Stanford Institute for Materials and Energy Sciences, SLAC National Accelerator Laboratory, Menlo Park, CA 94025, USA}
\affiliation{Department of Applied Physics, Stanford University, Stanford, CA 94305, USA}
\author{P. S. Kirchmann}
\affiliation{Stanford Institute for Materials and Energy Sciences, SLAC National Accelerator Laboratory, Menlo Park, CA 94025, USA}

\author{I. R. Fisher}
\author{Z. X. Shen}
\affiliation{Stanford Institute for Materials and Energy Sciences, SLAC National Accelerator Laboratory, Menlo Park, CA 94025, USA}
\affiliation{Department of Applied Physics, Stanford University, Stanford, CA 94305, USA}

\author{D. A. Reis}
\affiliation{Stanford PULSE Institute, SLAC National Accelerator Laboratory, Menlo Park, CA 94025, USA}
\affiliation{Stanford Institute for Materials and Energy Sciences, SLAC National Accelerator Laboratory, Menlo Park, CA 94025, USA}
\affiliation{Department of Applied Physics, Stanford University, Stanford, CA 94305, USA}
\affiliation{Department of Photon Science, Stanford University, Stanford, CA 94305, USA}

\begin{abstract}
We present a combined ultrafast optical pump-probe and ultrafast x-ray diffraction measurement of the CDW dynamics in SmTe$_3$ at 300~K. The ultrafast x-ray diffraction measurements, taken at the Linac Coherent Light Source reveal a $\sim 1.55$~THz mode that becomes overdamped with increasing fluence. We identify this oscillation with the lattice component of the amplitude mode. Furthermore, these data allow for a more clear identification of the frequencies present in the optical pump-probe data. In both, reflectivity and diffraction, we observe a crossover of the response from linear (for small displacements) to quadratic in the amplitude of the order parameter displacement. Finally, a time-dependent Ginzburg-Landau model captures the essential features of the experimental observations. 
\end{abstract}


\maketitle
\section{Introduction}

Charge density waves (CDWs)~\cite{gruner1988} are broken symmetry states of metals that spontaneously develop a valence charge modulation and a gap in the electronic structure concomitant with a frozen lattice distortion with a well-defined wavevector, $\mathbf{q}_{\rm cdw}$. The lattice exhibits a Kohn anomaly, a soft phonon mode of the symmetric phase, whose frequency $\omega(\mathbf{q}_{\rm cdw})$ decreases as the transition temperature, $T_c$, is approched from above.
In the original mechanism proposed by Peierls, the CDW forms due to an electronic instability that occurs because of Fermi-surface nesting between bands separated by $\mathbf{q}_{\rm cdw}$. Later arguments, however, showed that Fermi-surface nesting does not provide predictive power: in most 2D systems the CDW wavevector is not the optimum nesting wavevector, and the wavevector dependence of the electron-phonon matrix elements must be included to obtain the correct ordering wavevector~\cite{johannes2008}.  

Over the last few decades we have seen tremendous progress towards materials control at ultrafast timescales using light pulses~\cite{basov2017}. With the goal of understanding the materials dynamics, CDWs provide attractive model systems to study the dynamics of order parameters and fluctuations when driven out of equilibrium. 
In addition,  the CDW long-range order typically occurs at a well-defined wavevector and the transition can be modeled with a small number of degrees of freedom. 
Pump-probe methods have the ability to probe the system both near and far from equilibrium as the transition occurs and, from the dynamics, obtain information about the coupling between the participating degrees of freedom. 
Various ultrafast techniques have been used to probe the transient dynamics of charge density waves: ultrafast x-ray~\cite{mohr-vorobeva2011,huber2014,moore2016} and electron~\cite{eichberger2010} diffraction probed the structural transformation by measuring the intensity of the CDW Bragg peaks; ultrafast optical spectroscopy can probe the spectrum of low-energy excitations and their transient dynamics with unprecedented frequency resolution~\cite{demsar1999,yusupov2008,yusupov2010}, and time- and angle-resolved photoemission spectroscopy can probe the transient electronic gap and quasiparticle populations~\cite{schmitt2011,rettig2014,leuenberger2015}.

The rare-earth tri-tellurides ($R$Te$_3$ with $R$ a rare earth ion) has attracted much attention as a model system for studying the interplay between Fermi-surface nesting~\cite{brouet2004} and electron-phonon coupling~\cite{Eiter2013,maschek2015,moore2016} in CDW phenomena. 
Here we present ultrafast optical pump-probe and ultrafast x-ray diffraction on SmTe$_3$ at 300~K. SmTe$_3$ undergoes a CDW transition at $T_c = 416$~K. The high-symmetry phase of SmTe$_3$ crystallizes in the $Cmcm$ space group~\cite{norling1966} with lattice constants $a=4.333$, $b=25.68$, $c=4.336$~\AA. In the samples studied here, the long axis, $b$, is perpendicular to the sample surface. Below $T_c$ the material develops a static CDW~\cite{ru2008} with an incommensurate wavevector $\mathbf{q}_{\rm cdw} = (0,0,q) \approx (0,0,2/7)$ (reciprocal lattice units, rlu). 

Using ultrafast hard x-ray pulses from the Linac Coherent Light Source (LCLS), we measured the dynamics of the lattice component of the order parameter at $\mathbf{q}_{\rm cdw}$ at varying degrees of photoexcitation. 
Comparing the pump-probe reflectivity data with the x-ray results allows for the identification of the  features observed in reflectivity and separate the zone-center optical phonons from the relevant mode at $\mathbf{q}_{\rm cdw}$. 
We observe a clear crossover of the response, both in the reflectivity and diffraction efficiency, from linear to quadratic in the order parameter (the lattice displacement) as a function of the excitation density. 
This is manifested in diffraction as an oscillation of the intensity of the nearly suppressed CDW Bragg peaks, and corresponds to oscillations in the new potential energy surface of the symmetric phase without the static CDW order. 
Finally, a time-dependent Ginzburg-Landau model explains semi-quantitatively the dynamics of the lattice order over the range of fluences measured.

\section{Ultrafast X-ray diffraction}

\begin{figure}[htb]
\centering 
\includegraphics[width=0.95\columnwidth]{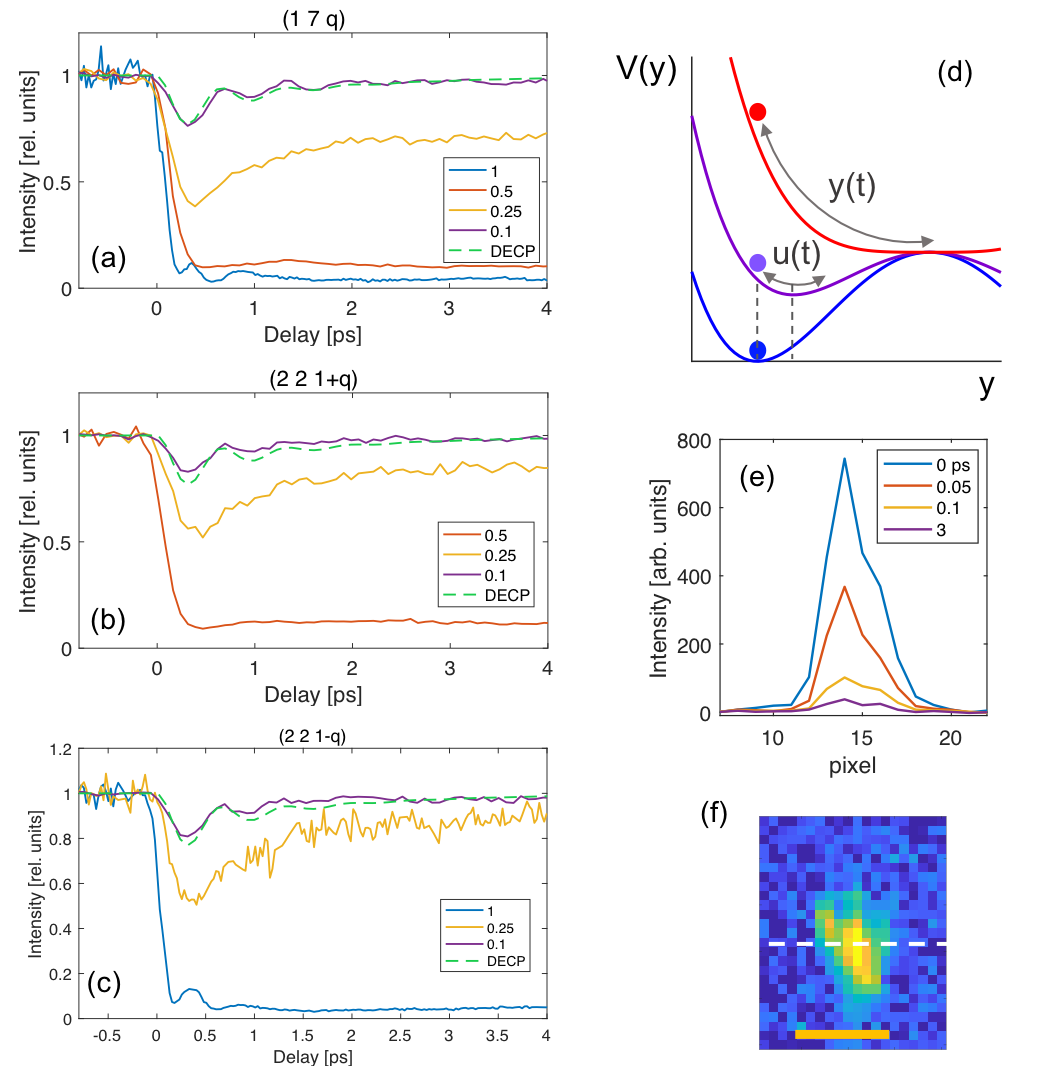} 
\caption[]{(color online) (a-c) Dynamics of the CDW X-ray diffraction for several incident fluences for CDW peaks (a) (1 7 q), (b) (2 2 1+q) and (c) (2 2 1-q).  The caption in (a-c) indicates the incident fluence in mJ/cm$^2$. The green dashed line is a fit using the DECP model from \cite{zeiger1992}, the parameters of the fit (given in the text) are the same for all panels. (d) Schematic of a symmetric double-well Ginzburg-Landau potential. The three curves represent three excitation levels. The view is expanded near the left minimum and each dot represent the initial displacement of the order parameter at $t=0$ for the different fluences. (e) lineout of the intensity across the detector along the dashed line indicated in (f). Panel (f) shows the detector image of the $(17q)$ CDW Bragg peak, the horizontal scale bars is $9\times 10^{-4}$~\AA$^{-1}$.
\label{fig:1} } 
\end{figure}

The ultrafast diffraction experiment was carried out at the XPP instrument at the LCLS~\cite{chollet2015} with x-ray pulses with $< 50$~fs in duration at a photon energy of 9.5 keV selected using a diamond double-crystal monochromator that provides $0.5$~eV bandwdith. The pump consisted of 50 fs pulses from a regenerative Ti:sapphire amplifier centered at 800 nm focused to a cross-sectional area of $0.05 \times 0.2$~mm$^2$. An area detector (CSPAD detector) was positioned at $\sim 1$~m from the sample and was rotated to capture the various Bragg reflections. To match the optical and x-ray penetration depth we implemented a grazing incidence geometry. The incidence x-ray angle of $0.3$ degrees was accurately calibrated by measuring the deflection by x-ray total external reflection at small angles.

In Fig.~\ref{fig:1} a-c we show the dynamics of the intensity of three different CDW diffraction peaks, $\tilde{I} = I(\mathbf{q}_{\rm cdw},t)/I_0(\mathbf{q}_{\rm cdw})$ for various incident fluences.  The intensity is integrated over the entire peak  and normalized by the intensity without the pump, $I_0(\mathbf{q}_{\rm cdw})$. For clarity in what follows we drop the $(\mathbf{q}_{\rm cdw})$ argument. The low fluence traces (top trace in each panel) show a $20\%$ decrease in $\tilde{I}$ and clear oscillations with frequency (period) $\sim 1.55$~THz (650~fs) that decay within a few ps. Since the scattering isolates the phonon wavevector, here only phonon modes with wavevector $\mathbf{q}_{\rm cdw}$  contribute to $I$. Thus, x-ray scattering at the CDW wavevector avoids the contribution from other Raman-active phonon modes at the zone-center~\cite{yusupov2008} and isolates the lattice component of the amplitude mode (AM) that only occurs at $\mathbf{q}_{\rm cdw}$. Therefore as we discuss later, we ascribe these oscillations to vibration of the lattice component of the amplitude mode.
As the pump fluence increases, we see a drastic decrease in $\tilde{I}$, which reaches complete suppression of the intensity for the highest fluence of $1$~mJ/cm$^2$. 


Since the CDW distortion is small compared with the lattice parameters the normalized diffraction efficiency is $\tilde{I} \propto y^2$, where $y = x/x_0$, $x$ is the order parameter and $x_0$ its equilibrium amplitude. Fig.~\ref{fig:1} (d) shows a schematic diagram of a Ginzburg-Landau potential as a function of the parameter $y$ with the three curves illustrating the ground state and two different excitation levels. For small lattice displacements from the equilibrium value (middle curve in Fig.~\ref{fig:1} (d)), $y \approx 1 - u(t)$, with $u$ the amplitude mode (AM) displacement, and $\tilde{I} \propto y^2 \approx 1 - 2 u(t)$.  As expected, the intensity is linear in the AM displacement for small $u$. 
For the dynamics in $u(t)$ we assume a model of displacive excitation of coherent phonons (DECP)~\cite{zeiger1992}. This model is equivalent to a sudden shift in the minimum of the potentials in Fig.~\ref{fig:1} (d) (bottom and middle curves), with approximately no change in the curvature. The following expression describes the small-amplitude oscillatory component of the photoexcited phonon as well as the change in its equilibrium position~\cite{zeiger1992}
\begin{equation}\label{eq:1}
	u(t) = A \{e^{-\beta t} - e^{-\gamma t} (\cos \Omega t - \frac{\beta'}{\Omega} \sin \Omega t ) \} \Theta(t),
\end{equation}
where $\Omega = \sqrt{\omega_0^2 - \gamma^2}$, $\omega_0$ is the bare frequency of the oscillator, $\beta' = \beta - \gamma$, $\gamma$ is the oscillation damping constant, $\beta$ is time-constant for the recovery of the shifted equilibrium position and $\Theta(t)$ is a step function. Here the first term corresponds to the non-oscillatory amplitude of the phonon that results from the shifted minimum of the potential and the second term results in the oscillations in the intensity. The green dashed traces in Fig.\ref{fig:1} (a-c) correspond a fit of Eq.~(\ref{eq:1}) to the lowest fluence trace with $\Omega/2 \pi = 1.55$~THz, $A = 0.085$, $\gamma = 1.8$~THz, and $\beta = 0.65$~THz.

As the fluence increases we observe a complete softening of the AM at $\sim 0.25 - 0.5$~mJ/cm$^2$.  At 1 mJ/cm$^2$ we observe two overdamped oscillations at higher frequency than those in the low fluence trace (see e.g. 1~mJ/cm$^2$ traces in Fig.~\ref{fig:1} (a) and (c)) and can be understood qualitatively as the order parameter crossing to the opposite side of the double well, as shown in the schematic diagram in Fig.~\ref{fig:1} (d) (top curve). 
At high excitation densities the displacements are large and the expansion of $y$ in terms of small displacements $u$ is not longer valid. In this regime, the expansion of $y$ about small displacements from the equilibrium does not hold and one must retain the full form $\tilde{I} \approx y(t)^2$. We point out that, because of the $\sim y^2$ dependence the period of the oscillation observed in the 1~mJ/cm$^2$ traces is half of the period of oscillation of $y(t)$ around $y=0$ in the new potential (top trace in Fig.~\ref{fig:1} (d)), as has been previously observed in K$_{0.3}$MoO$_3$\cite{huber2014}. 
This crossover from linear to quadratic as the fluence of the pump increases is also observed in the pump-probe reflectivity presented below (Fig.~\ref{fig:2}).  We note that the oscillation frequency in the low and high excitation regimes are not related since the curvature of the corresponding potentials (red and blue curves in Fig.~\ref{fig:1}) are not the same. 


\section{Ultrafast optical reflectivity}

As we see next, many of the features of the order parameter dynamics pointed out above are also visible in an ultrafast reflectivity probe. We present here an optical-pump, optical-probe reflectivity measurement of SmTe$_3$ for similar excitation fluences. The transient reflectivity at 800 nm was measured with 45~fs pulses from a Coherent RegA laser system at a repetition rate of 250~kHz. The pump and probe were near-collinear at normal incidence and the pump was chopped at 2~kHz. The reflected beam intensity was collected with a photodiode and the signal at the chopper frequency was measured with a lock-in amplifier. The pump and probe beam sizes (full-width at half maximum, FWHM) at the sample position were $60~\mu m$ and $25~\mu m$, respectively. 

\begin{figure}[htb]
\includegraphics[width=0.99\columnwidth]{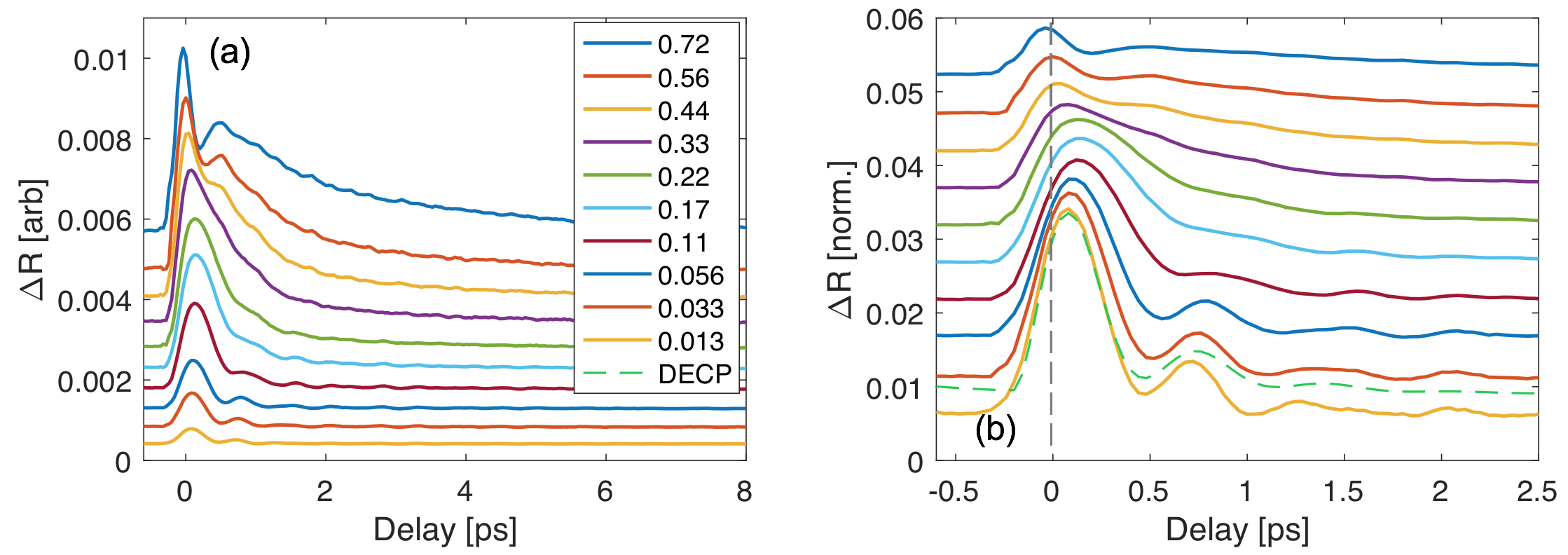} 
\caption[]{(color online) (a) Differential reflectivity at a wavelength of 800 nm as a function of pump-probe delay. The pump wavelength was 800nm and the incident fluence for each trace is labeled in the caption in mJ/cm$^2$. (b) Zoomed view of the data in (a) normalized by the incident fluence. The dashed curve is a DECP fit of the 0.033 mJ/cm$^2$ with $\Omega/2 \pi = 1.55$~THz, $\gamma = 2.5$~THz, $\beta = 1.95$~THz.
The vertical dashed line marks the $t=0$ point.  \label{fig:2} } 
\end{figure}

Fig.~\ref{fig:2}a shows the time-domain reflectivity of SmTe$_3$ at 300 K for increasing incident fluence, indicated in the caption (in mJ/cm$^2$). These data have more oscillatory components than the x-ray traces in Fig.~\ref{fig:1} because the reflectivity is modulated in principle by all possible Raman-active modes in the material consistent with selection rules. Fig~\ref{fig:2}b shows a zoomed view at early times of the same data in (a) but normalized by the respective fluence to highlight the features at low fluence. The dashed line in Fig.~\ref{fig:2}b shows a DECP fit to the trace for 0.033 mJ/cm$^2$ fluence, whose frequency most closely matches that of the 0.1 mJ/cm$^2$ x-ray data in Fig.~\ref{fig:1}. 
For low fluence the time-domain trace shows several oscillations corresponding to various Raman-active phonons, including the AM~\cite{yusupov2008,lavagnini2008a}, which seems to become overdamped as fluence reaches $F \sim 0.33$~mJ/cm$^2$. This is made clear when normalizing the time traces by the incidence fluence (Fig. \ref{fig:2} (b)), and is observed as a delay of the first maximum of oscillation. 
For $F \sim 0.7$~mJ/cm$^2$ (top trace in Fig.~\ref{fig:2} (a)) we observe a fast, single-cycle oscillation, whose period is shorter than that of the low fluence AM, and which resembles the high-fluence trace in the x-ray structure factor (compare with the high-fluence traces in Fig.~\ref{fig:1} (a) and (c)). The two have the same origin: due to symmetry both the diffraction intensity (the structure factor) and the dielectric function measured by reflectivity, must be quadratic in $y$ for the high-symmetry phase. Thus the dielectric function is $\epsilon(y) \approx \epsilon_0 + a y^2$. A similar argument as that made for explaining the diffraction holds: for small amplitudes of the AM, $y(t) = 1 - u(t)$ and the expansion of $\epsilon$ has a leading order term $\propto u(t)$, which means that the AM, $u$, can be Raman-active to first order in $u$. However, for large motions the expansion in $u$ is no longer valid but $\epsilon \propto y^2(t)$ for all range of values of $y$. 
This means that for large deviations of $y$ from the equilibrium ($y=1$), the motion is not probed through first-order Raman as in the case of the AM, even though both regimes are driven by the DECP mechanism. The fact that $\epsilon \propto y^2(t)$ means that the deviations of the order parameter from $y=0$ couple to the probe as a second order Raman process~\cite{ginzburg1980,yusupov2010,henighan2016}. This simple symmetry argument explains why the reflectivity and the structure factor behave similarly when reaching the $y=0$ symmetric point at high fluence.

\begin{figure}[htb]
\centering 
\includegraphics[width=0.8\columnwidth]{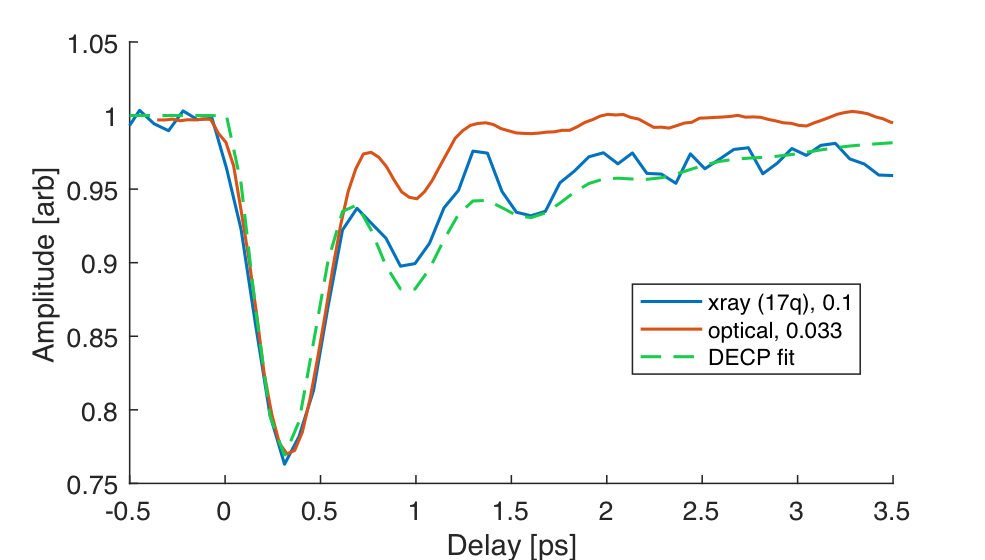} 
\caption[]{(color online) Comparison between x-ray and optical data for the lowest fluence. The dashed line is the DECP fit of the x-ray data in Fig.~\ref{fig:1}.
\label{fig:xray_optical_comp} } 
\end{figure}

For comparison between x-ray and optical results we plot in Fig.~\ref{fig:xray_optical_comp} the two traces of low fluence x-ray and optical data that most closely match (the optical data has been inverted and scaled to match the overall amplitude). The dashed line here is the DECP fit from Fig.~\ref{fig:1} (a). We observe that the oscillations in the 0.033 mJ/cm$^2$ optical reflectivity curve best match the low fluence oscillations in the x-ray data (0.1 mJ/cm$^2$), which provides a robust comparison between the fluences of the two measurements and removes systematic errors when comparing excitation levels between them. This comparison very clearly suggests that the soft mode component in the optical data is related to the lattice component of the order parameter at $\mathbf{q}_{\rm cdw}$.

\begin{figure}[htb]
\centering 
\includegraphics[width=0.9\columnwidth]{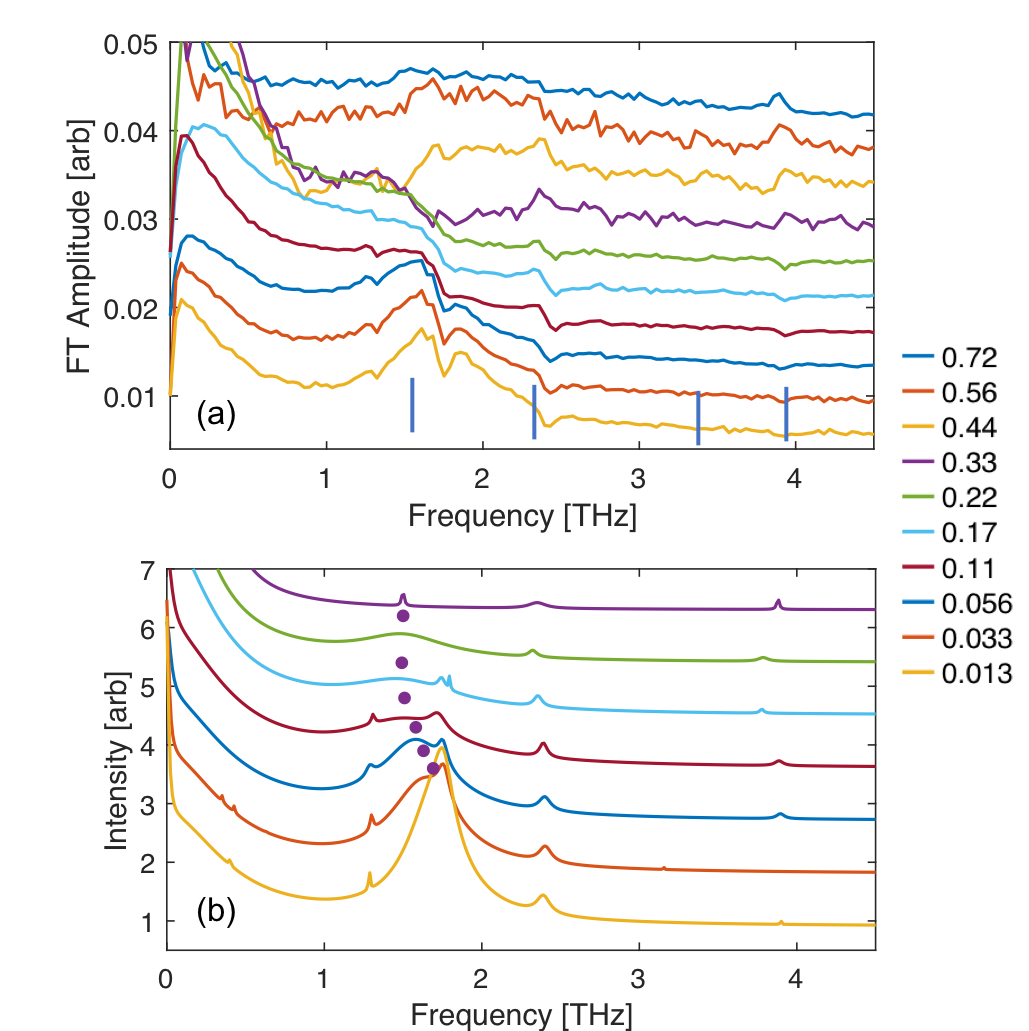} 
\caption[]{(color online) Frequency spectrum of the optical reflectivity data in Fig.~\ref{fig:2} extracted from (a) a Fourier-transform (FT) after background subtraction, and (b) the linear prediction procedure discussed in the text~\cite{barkhuijsen1985}.  The incident fluence for each trace is indicated next to the FT traces (in mJ/cm$^2$). The purple dots in (b) mark the frequency of the LP component ascribed to the soft-mode. Vertical bars indicate the frequencies of the most prominent phonon modes of SmTe$_3$ observed in Raman scattering~\cite{lavagnini2008a}.\label{fig::spectrum} } 
\end{figure}

Fig.~\ref{fig::spectrum} (a) shows the Fourier transform of the data in Fig.~\ref{fig:2} after subtraction of a double exponential that represents the non-oscillatory contribution from photoexcited quasiparticles~\cite{demsar1999,yusupov2010}. 
It is clear from Fig.~\ref{fig::spectrum} (a) that there are several modes in the data with a broad double feature at $\sim 1.65$~THz and two clear modes at $2.5$ and $3.95$~THz. The most prominent broad feature at $\sim 1.65$~THz softens as a function of fluence, while the frequency of the other modes remain static as fluence increases. The vertical bars indicate the frequencies of the most prominent Raman active modes of SmTe$_3$ observed in \cite{lavagnini2008a}. 

As an alternative approach to obtain the frequency content of these oscillations we treat the data using a linear prediction algorithm. This algorithm operates on the time-domain data directly, and can be thought of as a more robust way of fitting exponentially-decaying cosines to the time traces, in which the number of oscillators is determined from the statistical properties of the data~\cite{barkhuijsen1985,wise1987}. The procedure relies on linear prediction of each value in terms of previous ones, and thus yields a linear least squares fit of the exponentially-decaying cosines to the time traces, rather than a nonlinear fit that depends on the initial guess values of the parameters. As described in \cite{barkhuijsen1985} the procedure outputs the frequencies, decay constants, amplitudes and phases of the oscillators and can include pure decaying exponentials (zero frequency) components. For presentation purposes we compute the total spectrum as a sum of lorentzian functions with parameters given by the frequencies and damping constants output by the algorithm.
Fig.~\ref{fig::spectrum} (b) shows the spectrum of the low fluence traces obtained by applying linear prediction to the time-domain traces in Fig.~\ref{fig:2} (a). 
The frequency of the soft-mode obtained by this method is indicated by the dot $\sim 1.6$~THz above the corresponding trace. Note that not only does the frequency decrease but the width of this soft-mode component increases with increasing fluence, consistent with the observations in the FT (Fig.~\ref{fig::spectrum} (a)). 
We point out that the AM, whose frequency is $\sim 2.2$~THz near $10$~K~\cite{yusupov2008}, softens strongly as temperature increases towards the critical temperature, $T_c$, and crosses other phonon modes near 1.75~THz at 100~K below $T_c$\cite{yusupov2008}. Extrapolation from the literature observation for HoTe$_3$,  DyTe$_3$ and  TbTe$_3$\cite{yusupov2008} to SmTe$_3$ indicates that the AM crosses the 1.75~THz mode around $T \sim 280$~K. Thus the AM at 300~K is already significantly softened and photoexcitation likely contributes additional softening. 
We further note that the 2.5~THz mode does not soften and remains visible even for fluences $> 0.25$~mJ/cm$^2$ where the CDW diffraction is strongly suppressed. Taking into consideration the x-ray and optical comparison in Fig.~\ref{fig:xray_optical_comp}, we identify the soft-mode at $1.6$~THz with the $1.55$~THz oscillations in the low-fluence x-ray traces in Fig.~\ref{fig:1} and assign it to the  AM. Note that trARPES shows modes at 2.2~THz and 2.5~THz at relatively low-fluence ~\cite{leuenberger2015}, and per the discussion above, the 2.2~THz mode in ~\cite{leuenberger2015} is identified with the softened 1.6~THz mode observed here. 


\section{Time-dependent Ginzburg-Landau model}

We now turn our attention to the high fluence x-ray data. As the incident fluence, $F$, increases we observe a suppression of the oscillatory dynamics for fluences above $0.25$~mJ/cm$^2$, and almost complete extinction of the CDW intensity at $F > 0.5$~mJ/cm$^2$ (Fig.~\ref{fig:1}). We note that the Bragg peak width (inverse of the correlation length) does not appreciably change over the range of delays probed here, as seen in the Bragg peak cross sections in Fig.~\ref{fig:1} (e) and consistent with previous resonant diffraction measurements~\cite{moore2016}. This suggests that, unlike the thermal transition~\cite{ru2008} where the correlation length diverges, the ultrafast destruction of the CDW order proceeds without the creation of topological defects~\cite{wslee2012} at these timescales.

We model the dynamics using a time-dependent extension of the Ginzburg-Landau formalism for second order phase transitions~\cite{yusupov2010,huber2014}. The model assumes that the dynamics of the transition can be described by a real order parameter and ignores phase fluctuations at these timescales~\cite{yusupov2010}. We also assume that the electronic degrees of freedom follow adiabatically the dynamics of the lattice; under this assumption both the electronic and lattice degrees of freedom may be described by a single parameter~\cite{schaefer2010,schaefer2014}.

The Ginzburg-Landau (GL) potential is taken to be of the form

\begin{equation}\label{eq:gl_potential}
V(x)	 = \frac{1}{2}a (\eta-1) x^2 + \frac{1}{4} b x^4.
\end{equation}
Here $\eta \geq 0$ acts as a control parameter that in equilibrium is $\eta = T/T_c$. For $\eta < 1$ the system is in a double well configuration (lowest trace in Fig.~\ref{fig:1} (d)) with two minima at $x_0 = \pm\sqrt{a/b}$. In the small amplitude regime where $\eta \approx 0$ this model reduces to a DECP model with $a = \Omega^2/2$ and the low amplitude dynamics simplify to those of Eq.~(\ref{eq:1}).
For $\eta \ge 1$ the potential has a single minimum at $x=0$ (top trace in Fig.~\ref{fig:1} (d)). Equation (\ref{eq:gl_potential}) can be simplified further using the normalized order parameter, $y = x/x_0$. The equation of motion for $y$ substituting $a = \Omega^2/2$ is
\begin{equation}\label{eq:gl_motion}
	\frac{2}{\Omega^2}\ddot{y} + (\eta(t) - 1) y + y^3 + \frac{2\Gamma}{\Omega^2} \dot{y} = 0,
\end{equation}
with initial conditions $y(0) = 1$, $\dot{y}(0) = 0$. The last term accounts for damping of the dynamics and we assumed that we can describe the photoexcitation by introducing a time dependent $\eta = \eta(t) =e^{-t/\tau} \Theta(t)$ with  $\Theta(t)$ a unit step function. To account for the experimental observations, the relaxation rate of the photoexcited quasiparticles, $\tau$, is assumed to be fluence-dependent and is allowed to vary when fitting the model. This effect can be observed in the slow, non-oscillatory component of the dynamics in both the optical and x-ray data. 
In practice the values of $\eta$ and $\Gamma$ are determined by fitting the numerical solution of Eq.~(\ref{eq:gl_motion}), $y(t)$, to the intensity data in Fig.~\ref{fig:1}, assuming that the CDW intensity is $\tilde{I} \propto y^2(t)$. 

\begin{figure}[htb]
\centering 
\includegraphics[width=0.95\columnwidth]{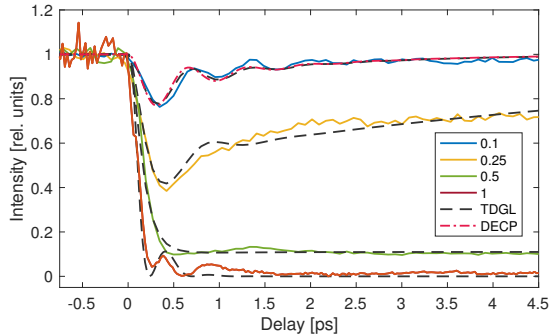} 
\caption[]{(color online) Time dependent Ginzburg-Landau model of the dynamics of the order parameter. The solid lines are the time dependence of the structure factor for the $(17q)$ reflection as in Fig.~\ref{fig:1} (a). The dark dashed lines are the solutions to the TDGL for the values reported in Table~\ref{table0}. For comparison, the dashed-dotted line shows the DECP fit from Fig.~\ref{fig:1}.
\label{fig:5} } 
\end{figure}

Figure \ref{fig:5} shows the fit of the time-dependent Ginzburg-Landau model (dashed lines) together with the fitted experimental data for the $(17q)$ Bragg peak from Fig.~\ref{fig:1} (a) for various fluences (solid lines). The frequency was only varied when fitting the 0.1 mJ/cm$^2$ data and was kept fixed at the resulting value when fitting the other fluences. The fit parameters are given in table \ref{table0}. Note that, except for $0.1$~mJ/cm$^2$, for the other fluences the magnitude of the fitted $\eta$ are within $ < 10\%$ of the expected amplitudes based on a linear scaling of $\eta$ by the corresponding fluence. As can be clearly seen in Fig.~\ref{fig:5}, this model reproduces the dynamics of the structure factor surprisingly well for the entire delay and fluence ranges. 
In the limit of low fluence, the solution $y(t)$ closely matches the DECP fit from Eq.~(\ref{eq:1}) as shown by the comparison with the dotted-dashed line in Fig.~\ref{fig:5}. At this fluence the model predicts a $\sim 20\%$ suppression of the intensity together with time-dependent oscillations due to the low amplitude vibrations of the AM as observed experimentally in the top trace of Fig.~\ref{fig:5}. The initial suppression in $\tilde{I}$ sharply increases with fluence and the motion in $y(t)$ seems to become overdamped at $F = 0.25$~mJ/cm$^2$ (which corresponds to $\eta = 0.5$, see Table~\ref{table0}). The critical point $\eta = 1$ is reached for $F \sim 0.5$~mJ/cm$^2$, which achieves nearly complete suppression of the CDW Bragg peak. 
At 1 mJ/cm$^2$ $\eta = 2$, and the system is pushed well into the high-symmetry phase where the potential has a single minimum at $y=0$ (top trace in Fig.~\ref{fig:1} (d)); after the sudden excitation the order parameter crosses the $y = 0$ point and performs two overdamped oscillations before fully decaying, as seen in Fig.~\ref{fig:1} (a) and (c). A similar crossover behavior is observed in the reflectivity data in Fig.~\ref{fig:2} for comparable fluences ($F > 0.5$ in Fig.~\ref{fig:2}). The current model provides a phenomenological description of these dynamics and explains the behavior of both the x-ray and reflectivity data over the entire regime of fluences.

\begin{table}[h]
\begin{tabular}{ l | c | c | c | r }
    F (mJ/cm$^2$) & $\Omega/(2\pi)$ (THz) & $\eta(0)$ (arb) & $\Gamma$ (ps$^{-1}$) & $\tau(F)$ (ps) \\
	\hline

  
  \hline			
  0.1& 1.6 & 0.16 & 3.78 & 1.63 \\
  0.25& 1.6$^*$ & 0.47 & 6.27 & 7.03 \\
  0.5& 1.6$^*$ & 0.89 & 9.07 & 1.5$\times 10^4$ \\
  1& 1.6$^*$ & 2.01 & 5.94 & $10^5$ \\
  \hline  
\end{tabular}
\caption[]{Parameters of the TDGL fit. Values labeled $^*$ were fixed at the result of the fit for 0.1~mJ/cm$^2$.\label{table0}}
\end{table}

\section{Conclusions}

In conclusion, we have presented a comprehensive ultrafast x-ray and optical study of the lattice dynamics of SmTe$_3$. We used x-ray diffraction to directly probe the lattice component of the order parameter at $\mathbf{q}_{\rm cdw}$ and isolate the AM from other Raman-active phonon modes that appear in reflectivity. In the high excitation regime, the average order parameter reaches the symmetric position and can overshoot for even higher fluence excitation. This effect is observed both in x-ray diffraction and optical reflectivity and is explained as a crossover from linear to quadratic dependence of both the structure factor and dielectric function on the order parameter. Finally, a time-dependent Ginzburg-Landau model describes the large-amplitude dynamics of the order parameter over the entire range of displacements. 

\section{Acknowledgements}

We thank Samuel Teitelbaum for enlightening discussions and a critical reading of the manuscript. Preliminary x-ray charaterization was performed at BL7-2 at the Stanford Synchrotron Radiation Lightsource (SSRL). MK, TH, MT, DL, PSK, ZXS, PG-G, IRF and DAR were supported by the U.S. Department of Energy, Office of Science, Office of Basic Energy Sciences through the Division of Materials Sciences and Engineering under Contract No. DE-AC02-76SF00515. Use of the LCLS and SSRL is supported by the U.S. Department of Energy, Office of Science, Office of Basic Energy Sciences under Contract No. DE-AC02-76SF00515. JNC was supported by the Volkswagen Foundation. Additional X-ray measurements were performed at BL3 of SACLA with the approval of the Japan Synchrotron Radiation Research Institute (JASRI) (Proposal No. 2016A8008). 



%


\end{document}